\documentclass[conference]{IEEEtran}
\IEEEoverridecommandlockouts
\usepackage{cite}
\usepackage{amsmath,amssymb,amsfonts}
\usepackage{algorithmic}
\usepackage{graphicx}
\usepackage{textcomp}
\usepackage{xcolor}
\usepackage{hyperref}
\def\BibTeX{{\rm B\kern-.05em{\sc i\kern-.025em b}\kern-.08em
    T\kern-.1667em\lower.7ex\hbox{E}\kern-.125emX}}

\newcommand\T{\rule{0pt}{2.4ex}}       
\newcommand\B{\rule[-1.0ex]{0pt}{0pt}} 

\begin{document}


\title{An investigation of the reconstruction capacity of stacked convolutional autoencoders for log-mel-spectrograms}

\author{\IEEEauthorblockN{Anastasia Natsiou}
\IEEEauthorblockA{\textit{Technological University of Dublin} \\
Dublin, Ireland \\
anastasia.natsiou@tudublin.ie}
\and
\IEEEauthorblockN{Luca Longo}
\IEEEauthorblockA{\textit{Technological University of Dublin} \\
Dublin, Ireland \\
luca.longo@tudublin.ie}
\and
\IEEEauthorblockN{Se\'{a}n O'Leary}
\IEEEauthorblockA{\textit{Technological University of Dublin} \\
Dublin, Ireland \\
sean.oleary@tudublin.ie}
}

\maketitle

\begin{abstract}

In audio processing applications, the generation of expressive sounds based on high-level representations demonstrates a high demand. These representations can be used to manipulate the timbre and influence the synthesis of creative instrumental notes. Modern algorithms, such as neural networks, have inspired the development of expressive synthesizers based on musical instrument timbre compression. Unsupervised deep learning methods can achieve audio compression by training the network to learn a mapping from waveforms or spectrograms to low-dimensional representations. This study investigates the use of stacked convolutional autoencoders for the compression of time-frequency audio representations for a variety of instruments for a single pitch. Further exploration of hyper-parameters and regularization techniques is demonstrated to enhance the performance of the initial design. In an unsupervised manner, the network is able to reconstruct a monophonic and harmonic sound based on latent representations. In addition, we introduce an evaluation metric to measure the similarity between the original and reconstructed samples. Evaluating a deep generative model for the synthesis of sound is a challenging task. Our approach is based on the accuracy of the generated frequencies as it presents a significant metric for the perception of harmonic sounds. This work is expected to accelerate future experiments on audio compression using neural autoencoders.

\end{abstract}


\section{Introduction}

Limited memory and the requirements for fast transmission of data influenced researchers and engineers to investigate methods that reduce the dimensionality of the data. Computational methods have previously been proposed to tackle the issue. \textit{Principle Component Analysis (PCA)} \cite{abdi_principal_2010} constitutes an algorithm that linearly projects every sample to a lower dimensional space, by extracting the most significant information as a set of new orthogonal variables.

More recently, the rise of deep learning methods presented the ability of the non-linear projection of data to a lower dimension. Unsupervised networks such as autoencoders (AE) \cite{hinton_reducing_2006} or Generative Adversarial Networks (GANs) \cite{goodfellow_generative_nodate} have demonstrated promising results in extracting information from relatively big and complex datasets into a latent space. Architectures like these were originally used to extract meaningful patterns from images \cite{radford_unsupervised_2016} or videos \cite{denton_unsupervised_nodate}. However, a few studies have also been proposed on speech \cite{hsu_learning_2017} or even music compression tasks \cite{colonel_improving_2017}.

In this work, we aim to investigate stacked convolutional neural autoencoders for reducing the dimensionality of time-frequency representations of harmonic sounds. An autoencoder (AE) is a type of neural architecture composed of an encoder and a decoder. The encoder projects the input representation to a lower dimensional space while the decoder attempts to restore the representation back to its original dimensions.

Convolutional neural networks belong to a sub-category of artificial neural networks (ANN) that aim to extract features from the input samples based on local pattern detection. The network is characterized by spatially local connections between nodes using a feature map that sweeps the input matrix. Additional techniques such as pooling layers can further decrease the dimensionality of the data.

This paper presents a design of convolutional autoencoders and their improvements using regularization methods. Using multiple instruments in a specific pitch, this work demonstrates an exploration of timbre compression. The network attempts to project the log-mel-spectrogram of monophonic and harmonic sounds to a lower dimensional space. This compressed high-level representation is used for musical instrument timbre synthesis.

The paper is organised as follows. In Section \ref{relatedwork} we review previous works of autoencoders for the compression of audio representations. Section \ref{methodology} includes details on autoencoders, convolutional networks, regularization techniques, and the mel-spectrogram while in Section \ref{experiments} we demonstrate the conducted experiments. In Section \ref{evaluation} we present the evaluation methods used and we introduce a novel evaluation metric. Finally, Section \ref{results} demonstrates the results and a discussion while Section \ref{conclusion} represents a brief conclusion.

\section{Related Work}
\label{relatedwork}

Neural autoencoders have been used in previous studies aiming to extract meaningful information from audio representations. In the majority of this research, neural networks were developed to compress magnitude short-time Fourier transform frames \cite{sarroff_musical_nodate, colonel_improving_2017, colonel_autoencoding_2020, roche_autoencoders_2019, colonel2020conditioning}. One of the first attempts was in \cite{sarroff_musical_nodate} where shallow feedforward neural autoencoders managed to decrease the original representation by $25\%$. Denoising techniques were also applied for additional regularization of the autoencoders. On the improvement of the previous work, Colonel et al \cite{colonel_improving_2017} developed a four-layer multilayer perceptron with Adam optimizer and additive bias using L1 and L2 for weight regularization. In both of the previous studies, they evaluated their methods of measuring the mean square error (MSE) between the original and generated frame of the magnitude spectrogram frame. Feedforward autoencoders were also applied in \cite{colonel_autoencoding_2020} for transforming a 2049 Fast Fourier Transform (FFT) frame to a latent space of 8 values. In order to improve the performance of the network, they used augmentation techniques with first and second-order differences of the magnitude spectrum as well as additional MFCCs. The architecture was evaluated by applying Inverse-STFT (ISTFT) on the generated frames developing also a deterministic method for reconstructing the phase. Furthermore, Colonel et al \cite{colonel2020conditioning} also conducted an investigation on activation functions for timbre synthesis using autoencoders.

More thorough research for the generation of magnitude STFT frames was conducted in \cite{roche_autoencoders_2019}. Shallow and deep feedforward networks as well as recurrent and variational autoencoders were compared to PCA. The study demonstrated that only deep autoencoders and LSTMs were able to adequately reconstruct the original samples. The model was evaluated subjectively using the MSE but also objectively by statistically analysing quality ratings. The waveform was synthesized by ISTFT with the original phase when there were no modifications in the latent space or using the Griffin Lim algorithm \cite{griffin_new_1985} when the phase needed to be estimated. 

Our method is heavily based on the baseline autoencoder developed in \cite{engel_neural_2017}. In this work, neural autoencoders were developed to extract meaningful information from a sound clip. They tried different input representations such as raw audio, the real and imaginary part of the STFT, or the log-magnitude part of the power spectra. The model was objectively evaluated using MSE between the original and generated spectrogram but also visually compared using \textit{rainbowgrams}, which are instantaneous frequency colored spectrograms.

\section{Methodology}
\label{methodology}

\begin{figure*}\centering
\includegraphics[width=0.8\linewidth]{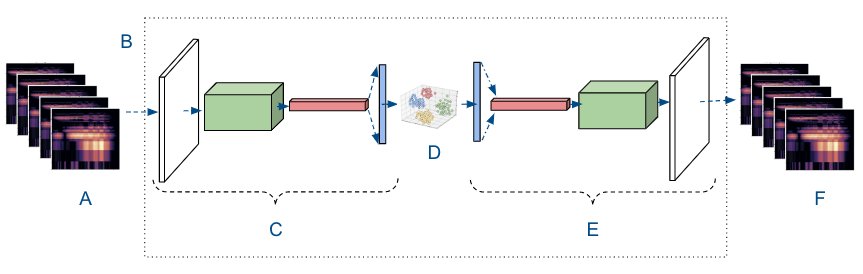}
\caption{Illustration of a stacked convolutional autoencoders architecture for the reconstruction of the log-mel-spectrogram. (A) Original log-mel-spectrogram. (B) Stacked convolutional autoencoder. (C) Encoder based on three convolutional layers. (D) Projection of the latent representation of multiple samples. (E) Decoder based on symmetric encoder. (F) Reconstructed log-mel-spectrogram.}
\label{stackedconvauto}
\end{figure*}

This research work is devoted to understanding the capacity of stacked convolutional autoencoders for the reconstruction of the log-mel-spectrogram. The design of the network is illustrated in Fig. \ref{stackedconvauto}, where an encoder based on convolutional networks aims to compress instrumental sounds to a lower dimensional space and a mirrored decoder attempts to reconstruct the samples from this high-level representation. We encourage the reader to listen to the synthesized sounds of the conducted experiments while moving forward with the paper. The waveform of the audio files \footnote{\url{https://anastasianat57.github.io/StackedConvolutionalAutoencoders}} is synthesized using the Griffith Lim algorithm \cite{griffin_signal_1984} as well as a sinusoidal signal reconstruction method \cite{natsiou_sinusoidal_2021}.

Towards the improvement of the design of convolutional autoencoders, regularization techniques are adopted to prevent overfitting and increase the performance of the network. In this research, we evaluate the effectiveness of autoencoders for sound synthesis purposes and we measure possible improvement using additional techniques that are expected to calibrate the model. This section presents a detailed explaination of autoencoders, convolutional neural networks and their components, a review of regularization methods, and the mel-spectrogram.

\subsection{Autoencoders}

An autoencoder is a neural architecture composed of an encoder and a decoder \cite{hinton_reducing_2006} trained together in an unsupervised manner. The encoder attempts to reduce the dimensionality of the original samples by extracting meaningful information in a non-linear procedure. Then, the decoder aims to reconstruct the original sample from the intermediate representation. This intermediate generated representation is called \textit{embedding} or \textit{latent representation}. 

For a feedforward single layer model, the encoder maps an input vector $x \in \mathbb{R}^d$ to an encoding $z \in \mathbb{R}^e$ where $d>e$ using a non-linear activation function $f(.)$  

\begin{equation}
    y = f(W x + b)
\end{equation}
where $W \in \mathbb{R}^{(e\times d)}$ represents the weights of the connections between the neurons and $b \in \mathbb{R}^e$ accounts for the bias term. The decoder maps $z$ back to the reconstructed $\widehat{x} \in \mathbb{R}^d$ using a similar approach
 
\begin{equation}
    \widehat{x} = f (W_{out}y + b_{out})
\end{equation}
where $W_{out} \in \mathbb{R}^{(d\times e)}$ and $b_{out} \in \mathbb{R}^d$.

To expand the initial architecture to a deep neural autoencoder, a similar approach is adopted. The encoder maps the input of a neuron $x_i$ to an output representation $x_j$ for $i,j<n$ as

\begin{equation}
    x \rightarrow x_1 \rightarrow x_2 \rightarrow ... \rightarrow x_n
\end{equation}
while the decoder reverses the previous procedure in a general way as

\begin{equation}
    x_n \rightarrow x_{n+1} \rightarrow x_{n+2} \rightarrow ... \rightarrow \widehat{x}
\end{equation}

The autoencoder is trained using a cost function that tries to minimize the error between the original and generated sample and updates the values of $W$ and $b$ using backpropagation. The activation and cost function used in autoencoders varies depending on the application and the training data.

\subsection{Convolutional Networks}

Convolutional neural networks (CNNs) consitute a regularized version of multilayer perceptrons. Their regularization depends on the fact that while feedforward neural networks investigate relationships within the whole sample, convolutional networks search for patterns locally. Their architecture is composed by three main components named convolutional layers, pooling, and fully connected layers.

\subsubsection{Convolutional Layers}

The convolutional layer is the most significant block of the CNNs. It is composed of a set of filters with smaller dimensions than the training samples and a set of learnable parameters. The filters are applied across the input data and the result creates the activation map. A few parameters define the convolutional layer, such as the number of filters, the number of moves over the input data (\textit{stride}) and \textit{zero-padding} which is the process of adding zeros outside the input matrix in order for the filters to fit the input size.

\subsubsection{Pooling}

Optionally, after every convolutional layer, a pooling layer is following the convolved output. Pooling is a dimensionality reduction process where a max or average filter matrix is applied across the convolved output. Although after pooling some information is lost, this technique can improve efficiency, reduce complexity, and limit the risk of overfitting.

\subsubsection{Fully Connected Layer}

In opposition to convolutional layers, fully connected layers directly connect each node of the output to the previous layer. This layer can be used to perform classification based on the features extracted from the convolutional layers and to manipulate the dimensionality of the outcome.

\subsection{Regularization Techniques}

One of the most considerable issues with deep neural networks is that they tend to overfit a training dataset. An overfitted model is a deep learning model that explicitly mimics the training samples by taking noise into consideration. These models are formulated with more parameters than necessary and become less accurate in predicting new data. To prevent overfitting, many techniques have been proposed including \textit{dropout} \cite{srivastava_dropout_nodate}, \textit{L1 and L2 regularization} \cite{girosi_regularization_1995}, data augmentation \cite{wong_understanding_2016}, or \textit{early stopping} \cite{orr_neural_1998}.

\subsubsection{Dropout}

A technique where in every iteration, the network selects some random nodes and excludes them from the learning process \cite{srivastava_dropout_nodate}. Therefore, in each iteration a different set of nodes predicts the output. Dropout can also be considered as an ensemble technique since numerous sub-networks are created for each learning step.

\subsubsection{L1 and L2 Regularization}

They constitute methods where a regularization term is added to the cost function of the network. This additional term produces an additional residual to the loss function and therefore overfitting is reduced \cite{girosi_regularization_1995}. L1 regularization (also known as \textit{Lasso}) updates the cost function using the Least Absolute Deviations (LAD) of the weight parameters as:

\begin{equation}
    Cost function = Loss + \frac{\lambda}{2m}\sum\Vert W \Vert
\end{equation}
where $\lambda$ corresponds to the regularization parameter and $m$ to the number of parameters. In L2 regularization the cost function is updated with the Least Square Errors (LSE) of the weight matrix:

\begin{equation}
    Cost function = Loss + \frac{\lambda}{2m}\sum\Vert W \Vert^2
\end{equation}

Regularization can be applied only on the weights $W$ (\textit{kernel regularization}), or only on the bias  term $b$ (\textit{bias regularization}), or on the output layer $y = Wx+b$ (\textit{activity regularization}).

\subsubsection{Data Augmentation}
An attempt to increase the training dataset by alternating the original samples. These additional data create extra noise to the network which can be used to prevent overfitting. In image processing, data augmentation can be achieved by flipping, scaling, or shifting the original images \cite{wong_understanding_2016} while in audio processing, common techniques include noise injection, time shifting, or speed alternation \cite{ko_audio_2015}.

\subsubsection{Early Stopping}

A validation dataset is used to calculate the loss function after each epoch. If the performance of the network in the training set increases while the performance in the validation set does not improve, then the model starts overfitting. Early stopping is used to prevent the network from mimicking the pattern of the training data \cite{orr_neural_1998}. \textit{Patience} denotes the number of epochs with no further improvement after which the training will be stopped.

\subsection{Mel-Spectrogram}

The spectrogram constitutes the time-frequency representation obtained by the application of the Fourier transform in overlapping fragments of time in an audio sample. The mel-spectrogram demonstrates a compressed form of the spectrogram based on human perception. The development of this mel-scale is based on the observation that the human ear demonstrates greater resolution at lower frequencies \cite{volkmann_scale_nodate}. The association between the frequencies in hertz $f$ and the mel-scale $mel$ (mel from 'melody') is shown in Eq.\ref{eq_mel}.

\begin{equation}
    mel = 2595log_{10}(1+\frac{f}{700})
    \label{eq_mel}
\end{equation}

The mel-spectrogram captures the most significant properties of the sound in a compressed form. Therefore, this time-frequency representation has been proven beneficial for deep learning applications since the memory and power requirements are reduced.

\section{Experiments}
\label{experiments}
\subsection{Dataset}

For the conducted experiments, we used a subsample of the NSynth dataset\footnote{\url{https://magenta.tensorflow.org/datasets/nsynth}}, which is a dataset of four-second monophonic notes. The subsample is composed of 3750 samples from a variety of instruments: guitar, bass, brass, synth, keyboard, flute, organ, mallet, vocal, reed, and string for a single pitch. The sounds were acoustic, electronic, or synthetic and could belong in different categories as per their velocity or acoustic quality.

The data preprocessing included the computation of the log-mel-spectrogram using a Blackman window of 690 samples, an FFT window of 1024 and 128 mel filter bands. The log-mel-spectrograms were later normalized to be transformed into the range $[0,1]$. The dataset was split into training, validation, and testing as 80/10/10.

\subsection{Models}

The proposed methodology demonstrates a stacked convolutional autoencoder with a mirrored encoder and decoder, like the one presented in Fig. \ref{stackedconvauto}. The two components are composed of three 2D convolutional layers with a kernel size of 4, stride of 2 and same padding. After experimental research, the hyperbolic tangent was used as an activation function for the convolutional layers while the softmax function was applied to the output layer to form the generated log-mel-spectrogram. Further investigation was conducted around kernel dimensions, filters, and pooling techniques. Experiments with additional regularization methods were also conducted, adopting techniques such as early stopping, dropout, and L1 and L2 kernel regularization and activity regularization. The network was trained using the ADAM optimizer \cite{kingma_adam_2017} with an initial learning rate of 0.001, and a mean square error loss function in batches of size 64. An early stopping patience limit
was set equal to 10 to avoid wasting resources during training. The experiments were conducted on a Tesla P100 GPU using the TensorFlow library\footnote{\url{https://www.tensorflow.org/}}.

\section{Evaluation}
\label{evaluation}

\begin{figure*}\centering
\includegraphics[width=0.9\linewidth]{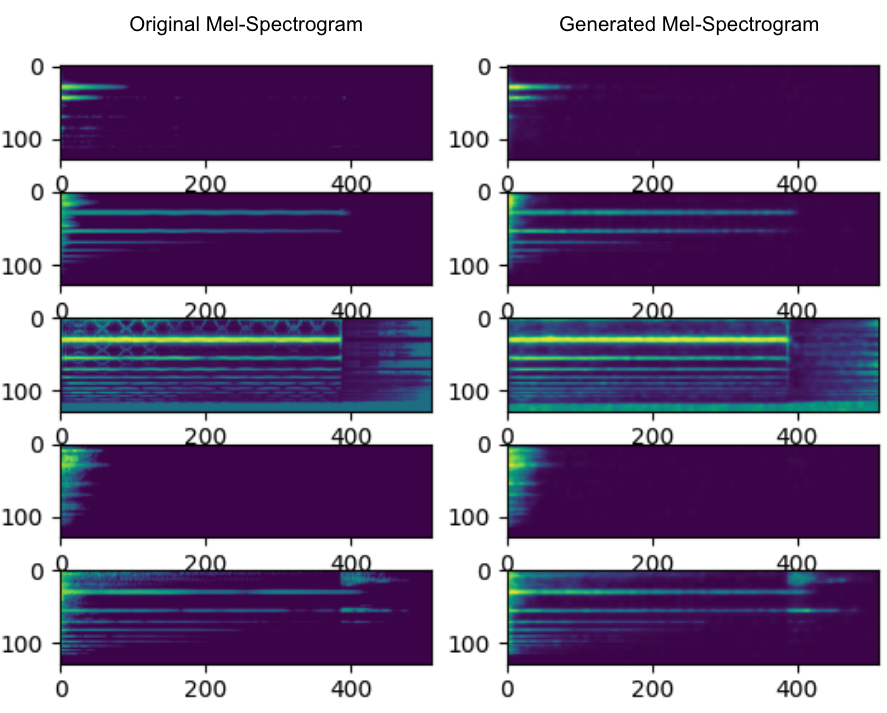}
\caption{Original and generated log-mel-spectrogram}
\label{specregeneration}
\end{figure*}

To evaluate the effectiveness of a generative network, many methods have been proposed. In these experiments, the root mean squared error (RMSE) between the original and generated spectrogram has been used. Another metric applied to measure the accuracy of the generated spectrogram was the structural similarity index (SSIM) which also demonstrated as a sufficient initial indicator for the comparison between the two spectrograms. However, these two objective metrics cannot accurately capture the quality of the spectrogram. A difference in any value of the spectrogram will produce the same mean square error regardless of the position of the value but the synthesized sound could be nonidentical. Furthermore, as it is depicted in Fig. \ref{specregeneration}, the majority of the spectrogram values are zero and therefore a small error can demonstrate dissimilarities between the two spectrograms. 

In order to evaluate the generated spectrograms, we developed a metric that highlights the significance of harmonics in timbre synthesis. Our approach is based on a peak detection algorithm to estimate the frequencies of the original spectrogram (\textit{OrigFreq}) and the frequencies of the generated spectrogram (\textit{GenFreq}) for every frame. Following, the identical frequencies (\textit{IdFreq}) constitute the retrieved frequencies that matched the relevant ones within a threshold of $\pm3\%$. This threshold can ensure that the detected peaks point to the same harmonic. In addition, to eliminate potential detected noise, harmonics with 30db from the peak amplitude are not taken into consideration.

The calculation of the original, generated and identical frequencies can then be used to evaluate the predicted samples using a precision-recall schema \cite{davis_relationship_2006}. Precision is computed as the fraction of the true positive values which are represented by the identical frequencies among the predicted frequencies as it is depicted in Eq. \ref{precision}.

\begin{equation}
    Precision = \frac{IdFreq}{GenFreq}
    \label{precision}
\end{equation}

Recall is the number of identical frequencies among the number of frequencies estimated in the original spectrogram. Recall, which can also be referred to as sensitivity, is demonstrated in Eq. \ref{recall}.

\begin{equation}
    Recall = \frac{IdFreq}{OrigFreq}
    \label{recall}
\end{equation}

Precision and recall are metrics to compute the relevance between the retrieved and relevant values. These two errors can later be combined into a single measurement called \textit{F1\_score} according to Eq. \ref{total_score}. F1\_score demonstrates the harmonic mean of precision and recall indicating a more accurate metric for the overall system.

 \begin{equation}
    F1\_score = \frac{2 \cdot Precision \cdot Recall}{Precision+Recall}
    \label{total_score}
\end{equation}

Precision, recall and F1\_score can score between 0 and 1, with higher value presenting a better performance. A high precision and low recall indicates a generated sample with less detected harmonics while a low precision and high recall indicates additional noise in the synthesized spectrogram.

\begin{figure*}\centering
\includegraphics[width=0.9\linewidth]{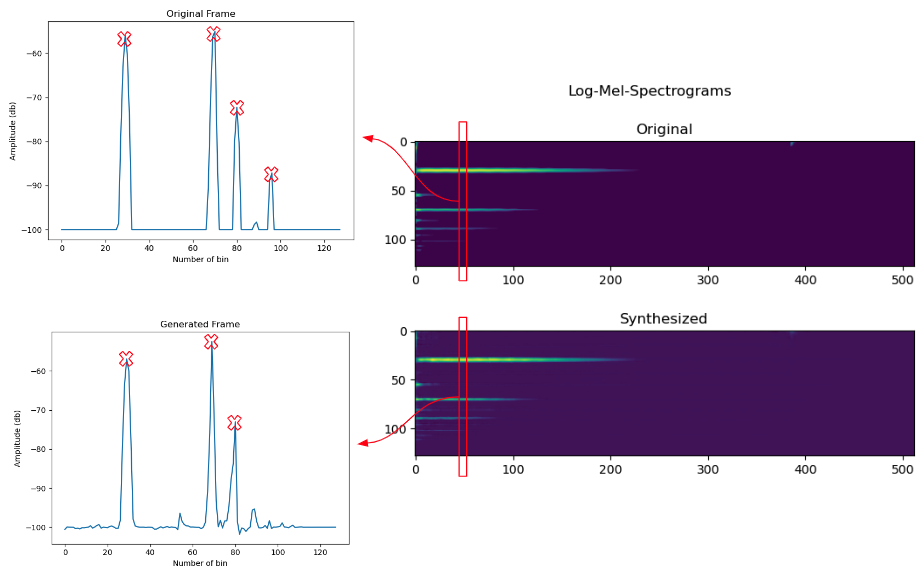}
\caption{Peak detection between the original and generated spectrogram}
\label{peakevaluation}
\end{figure*}

\section{Results and Discussion}
\label{results}

This research work’s main objective is to provide insight into the synthesis of timbre using stacked convolutional autoencoders. Additional parametrization and regularizations methods aim to minimize the error between the original and predicted audio samples. Each section of this paragraph provides results for an experimental set of parameters. Initially, we present the baseline autoencoder and then we discuss about experiments using regularization methods, pooling, dimensionality of the latent space and convolutional networks without a fully connected layer.

\subsection{Baseline}

The baseline design constitutes a neural autoencoder with three convolutional layers followed by a max pooling layer, and a fully connected layer. The encoder produces a latent representation of dimension 8192 and the network is trained with no other regularization methods than early stopping. The original and generated log-mel-spectrogram are illustrated in Fig. \ref{specregeneration}. As one can see, the reconstructed spectrogram is able to capture the general form of the original samples \footnote{\url{https://anastasianat57.github.io/StackedConvolutionalAutoencoders}}. 


\subsection{Regularization}

\begin{table}
\begin{center}
\begin{tabular}{|| c | c | c | c | c | c ||} 
 \hline
 Experiment & RMSE & SSIM & Recall & Precision & F1 \T\B \\ 
 \hline\hline
 No Reg	&	0.087	&	0.733	&	0.777	&	0.786	&	0.772	\T\B\\ 
 \hline
 Dropout E	&	0.097	&	0.686	&	0.707	&	0.797	&	0.732	\T\B\\
 \hline
 Dropout D	&	0.087	&	0.640	&	\textbf{0.832}	&	0.738	&	0.769	\T\B\\
 \hline
 Dropout ED	&	0.102	&	0.548	&	0.765	&	0.753	&	0.740	\T\B\\
 \hline
 KR L1 E	&	0.102	&	0.631	&	0.762	&	0.752	&	0.743	\T\B\\
 \hline
 KR L1 D	&	0.079	&	0.757	&	0.820	&	0.804	&	0.804	\T\B\\
 \hline
 KR L1 ED	&	0.080	&	0.761	&	0.829	&	0.794	&	0.804	\T\B\\
 \hline
 KR L2 E	&	0.086	&	0.722	&	0.804	&	0.791	&	0.790	\T\B\\ 
 \hline
 KR L2 D	&	\textbf{0.077}	&	\textbf{0.782}	&	0.819	&	0.829	&	\textbf{0.817}	\T\B\\
 \hline
 KR L2 ED&	0.079	&	0.766	&	0.793	&	\textbf{0.830}	&	0.801	\T\B\\
 \hline
 AR L1 E	&	0.085	&	0.778	&	0.788	&	0.826	&	0.798	\T\B\\
 \hline
 AR L1 D&	0.244	&	0.214	&	0	&	0	&	0	\T\B\\
 \hline
 AR L1 ED	&	0.244	&	0.215	&	0	&	0	&	0	\T\B\\
 \hline
 AR L2 E	&	0.162	&	0.395	&	0.541	&	0.697	&	0.552	\T\B\\
 \hline
 AR L2 D	&	0.244	&	0.177	&	0	&	0	&	0 \T\B	\\
 \hline
 AR L2 ED	&	0.238	&	0.210	&	0	&	0	&	0	\T\B\\
 \hline
\end{tabular}
\end{center}

\caption{Regularization techniques: A comparison between a network without regularization methods to a network with dropout on encoder (E), decoder (D), or encoder and decoder (ED), and to kernel regularizers (KR) and activity regularizers (AR)}
\label{table:regularization}
\end{table}

Additional enhancements were investigated in order to improve the performance of the baseline network. Table \ref{table:regularization} demonstrates an experimentation with different regularization techniques. More specifically, regularization methods are applied in order to improve the effectiveness of the autoencoder for the reconstruction of the log-mel-spectrogram. Dropout with a probability of 0.3 was applied only to the encoder (E), only on the decoder (D), or on both encoder and decoder (ED). Furthermore, experiments on kernel regularizers (KR) and activity regularizers (AR) using L1 or L2 regularization on encoder, decoder or both networks were conducted. 

The results showed that dropout did not improve the performance of the network. Specifically, dropout only on the decoder presents the same RMSE as the baseline but a lower F1\_score. However, an increased recall and decreased precision indicates that the model produces additional noise in the generated samples. Furthermore, dropout on the encoder or both encoder and decoder significantly deteriorated the performance of the network.

Using kernel regularizers as regularization methods, an improvement can be identified. Either L1 or L2 regularization function on the decoder produces the most accurate results. In particular, an L2 kernel regularizer on the decoder demonstrated the highest F1\_score and SSIM, and the lowest RMSE. The results are characterized by high precision and recall, and therefore generate the most promising spectrograms. Kernel regularizers on the encoder, or on both encoder and decoder also provide a slight improvement. Finally, activity regularizers spoiled the power of the neural networks yielding to a non-harmonic outcome. In most cases, the generated sounds produced are characterized by simple noise. 

Overall, stacked convolutional autoencoders can synthesize musical instrument timbre from a low dimensional representation. In some cases, regularization techniques such as L1 or L2 kernel regularization are able to enhance the performance of the network producing more accurate spectrograms. However, other regularization techniques, such as dropout, do not present any improvement and in even more extreme cases, like activity regularizers, the decoder generates random noise.

\subsection{Pooling}

\begin{table}
\begin{center}
\begin{tabular}{||c | c | c | c | c | c ||} 
 \hline
 Experiment & RMSE & SSIM & Recall & Precision & F1 \T\B \\ 
 \hline\hline
 Max Pooling	&	\textbf{0.087}	&	\textbf{0.733}	&	0.777	&	\textbf{0.786}	&	\textbf{0.772}	\T\B\\ 
 \hline
 Average Pooling	&	0.089	&	0.708	&	\textbf{0.787}	&	0.770	&	0.770	\T\B\\
 \hline
 No Pooling	&	0.106	&	0.676	&	0.692	&	0.732	&	0.692	\T\B\\
   \hline
\end{tabular}
\end{center}
\caption{A comparison between convolutional networks with max pooling, average pooling and no pooling}
\label{table:pooling}
\end{table}

The next set of experiments is based on the parametrization of the convolutional networks. More specifically, an investigation on pooling layers was conducted. Initially, max or average pooling layers were adopted after each convolutional layer of both the encoder and decoder. Lastly, a last experiment without pooling layers was also studied. In order for the network with no pooling layers to match the latent dimensions of the network with pooling, three additional convolutional layers were added. Table \ref{table:pooling} demonstrates the results from experiments based on pooling techniques. Based on the root mean square error, the structural similarity index and the F1\_score, networks with max or average pooling present similar performance. Although the F1\_score is almost identical, average pooling demonstrates a higher recall and a lower precision, indicating samples with additional noise. Finally, results from networks without pooling layers presented synthetic sounds of lower quality. Based on the results above, we can conclude that stacked convolutional autoencoders with a rough pooling approach, such as max pooling, can generate a more accurate audio time-frequency representation from a compressed low dimensional space of instrumental pitched sound.

\subsection{Dimensionality of the latent space}

\begin{table}
\begin{center}
\begin{tabular}{|| c | c | c | c | c | c ||} 
 \hline
 Experiment & RMSE & SSIM & Recall & Precision & F1 \T\B \\ 
 \hline\hline
 Latent Dim 8192	&	0.087	&	0.733	&	0.777	&	0.786	&	0.772	\T\B\\ 
 \hline
 Latent Dim 4096	&	\textbf{0.077}	&	0.796	&	\textbf{0.813}	&	0.828	&	0.814	\T\B\\
 \hline
 Latent Dim 2048	&	0.078	&	\textbf{0.798}	&	0.810	&	\textbf{0.837}	& \textbf{0.817}	\T\B\\
 \hline
\end{tabular}
\end{center}
\caption{Experiments on further compression of the latent space}
\label{table:latentdim}
\end{table}

To evaluate the effectiveness of neural autoencoders regarding their compression ability, an experimentation on the latent space dimension was conducted. Additional downsampling was studied by creating a latent representation of 4096 and 2048. The results are presented in Table \ref{table:latentdim}. According to the metrics, a more compressed representation is able to generate more accurate log-mel-spectrograms. Although the lower latent dimension was expected to reduce the performance of the network, the experiments revealed that a smaller latent space is able to synthesize spectrograms with a more smooth distribution of the higher partials leading to more natural sounds.

\subsection{Without a fully connected layer}

In the final set of experiments, we attempted to remove the dense layer of the autoencoder. A dense layer is necessary to manipulate the dimensions of the latent space and apply further compression. However, Table \ref{table:DenseNoDense} demonstrate that convolutional autoencoders without a dense layer generates spectrograms of better quality. That could be explained because the network preserves spatial information of the original samples.

\begin{table}
\begin{center}
\begin{tabular}{|| c | c | c | c | c | c ||} 
 \hline
 Experiment & RMSE & SSIM & Recall & Precision & F1 \T\B \\ 
 \hline\hline
 Dense	&	0.087	&	0.733	&	0.777	&	0.786	&	0.772	\T\B\\ 
 \hline
 No Dense	&	\textbf{0.063}	&	\textbf{0.861}	&	\textbf{0.856}	&	\textbf{0.859}	&	\textbf{0.855}	\T\B\\
 \hline
\end{tabular}
\end{center}
\caption{Network without a fully connected layer at the end of the convolutional blocks}
\label{table:DenseNoDense}
\end{table}

\subsection{Discussion}

From the above experiments, one can identify that stacked convolutional autoencoders are able to compress and reconstruct spectral representations of sound with adequate accuracy. As it is illustrated in Fig. \ref{specregeneration}, most of the frequency bands in the log-mel-spectrogram have zero values and it can be debatable whether the whole representation can be considered as the actual encoded information. However, hearing is a sense that demonstrates sensitivity and a slight change in the spectrogram can synthesize a significantly different sound. Therefore, the network needs to be exceptionally precise in the position and amplitude of the non-zero values.

Comparing the results with the state-of-the-art \cite{engel_neural_2017}, our sounds demonstrate phase coherence, making the sounds more pleasant. However, training the autoencoder in the whole NSynth dataset conditioned by the pitch can expand the variety of the produced sounds but also creates an additional perplexity which can reduce the performance of the network. Furthermore, using Wavenet \cite{oord_wavenet_2016} as a vocoder may not preserve the phase continuity in the waveform and therefore generate more noisy instrumental sounds.


\section{Conclusion}
\label{conclusion}

Deep generative models have been previously studied to synthesize timbre from a compressed representation. Most of these studies generate sound from frames of magnitude short-time Fourier transform. For this purpose, recurrent or fully-connected autoencoders have been proposed. In this work, we demonstrated an investigation on stacked convolutional autoencoders for the reconstruction of a spectrogram based on the perception of hearing and experimented with parametrization techniques and regularizations of the autoencoders. Moreover, we presented an evaluation method for calculating the accuracy of predicted frequencies in monophonic and harmonic musical sounds. The conducted experiments revealed that a latent representation compressed to 3\% of the size of the original data can synthesize timbre with adequate accuracy. Also, from a variety of regularization techniques applied to the network, some of them demonstrated improvement compared to the baseline autoencoder while others limited the abilities of the neural networks. The conducted research is expected to accelerate future studies on the compression of timbre and the evaluation of generated harmonic sounds. 

Future research work will expand the capabilities of the autoencoders including more diverse sounds in terms of pitch and instrument. An attempt for further compression will be studied by conditioning the network on additional temporal or spectral information. Moreover, incorporating temporal layers such as GRUs or LSTMs at the end of the encoder section is expected to influence the performance of the network. Finally, future work will outline an experimentation of the latent space using more perceptual properties. Following the previous work of Esling et al \cite{eslinggenerative} and Roche et al \cite{roche2021make}, an investigation around timbre descriptors will be conducted to enhance the structure of the latent representation generated by autoencoders and increase the accuracy of the synthesized audio. Numerous techniques of numeric or visual explanations can be adopted for the interpretation and evaluation of the generated latent space \cite{vilone2021classification, vilone2021notions}.

\section*{Acknowledgment}

This work was funded by Science Foundation Ireland and its Centre for Research Training in Machine Learning (18/CRT/6183).

\bibliographystyle{ieeetr}
\bibliography{ref}
\end{document}